\def\year{2021}\relax
\documentclass[letterpaper]{article} 
\usepackage{aaai21}  
\usepackage{times}  
\usepackage{helvet} 
\usepackage{courier}  
\usepackage[hyphens]{url}  
\usepackage{graphicx} 
\def\UrlFont{\rm}  
\usepackage{natbib}  
\usepackage{caption} 
\usepackage{amsmath}
\usepackage{amssymb}
\usepackage{bm}
\usepackage{amsthm}
\usepackage{algorithm}
\usepackage{algorithmicx}
\usepackage{algpseudocode}
\usepackage{subcaption}
\usepackage{booktabs}
\DeclareMathOperator*{\E}{\mathbb{E}}
\usepackage{mathtools}
\newtheorem{theorem}{Theorem} 

\DeclarePairedDelimiterX{\infdivx}[2]{(}{)}{%
	#1\;\delimsize\|\;#2%
}

\title{Tracking disease outbreaks from sparse data with Bayesian inference}

\author {
	Bryan Wilder,\textsuperscript{\rm 1}
	Michael J.\ Mina\textsuperscript{\rm 2},
	Milind Tambe\textsuperscript{\rm 1} \\
}
\affiliations {
	\textsuperscript{\rm 1} John A.\ Paulson School of Engineering and Applied Sciences, Harvard University  \\
	\textsuperscript{\rm 2} T.H.\ Chan School of Public Health, Harvard University \\
	bwilder@g.harvard.edu, mmina@hsph.harvard.edu, milind\_tambe@harvard.edu
}
\begin{document}
	
	\maketitle
	
	\begin{abstract}
		The COVID-19 pandemic provides new motivation for a classic problem in epidemiology: estimating the empirical rate of transmission during an outbreak (formally, the time-varying reproduction number) from case counts. While standard methods exist, they work best at coarse-grained national or state scales with abundant data, and struggle to accommodate the partial observability and sparse data common at finer scales (e.g., individual schools or towns). For example, case counts may be sparse when only a small fraction of infections are caught by a testing program. Or, whether an infected individual tests positive may depend on the kind of test and the point in time when they are tested. We propose a Bayesian framework which accommodates partial observability in a principled manner. Our model places a Gaussian process prior over the unknown reproduction number at each time step and models observations sampled from the distribution of a specific testing program. For example, our framework can accommodate a variety of kinds of tests (viral RNA, antibody, antigen, etc.) and sampling schemes (e.g., longitudinal or cross-sectional screening). Inference in this framework is complicated by the presence of tens or hundreds of thousands of discrete latent variables. To address this challenge, we propose an efficient stochastic variational inference method which relies on a novel gradient estimator for the variational objective. Experimental results for an example motivated by COVID-19 show that our method produces an accurate and well-calibrated posterior, while standard methods for estimating the reproduction number can fail badly.  
	\end{abstract}

	\section{Introduction}
	
	A key goal for public health is effective surveillance and tracking of infectious disease outbreaks. This work is motivated in particular by the COVID-19 pandemic but the methods we describe are applicable to other diseases as well. A central question is to estimate the empirical rate of transmission over time, often formalized via the reproduction number $R_t, t = 1...T$. $R_t$ describes the expected number of secondary infections caused by someone infected at time $t$. Accurate estimates of $R_t$ are critical to detect emerging outbreaks, forecast future cases, and measure the impact of interventions imposed to limit spread. 
	
	$R_t$ is typically estimated using daily case counts, i.e., the number of new infections detected via testing each day. Standard methods, including prominent dashboards developed for COVID-19, provide accurate estimates under idealized conditions for the observation of cases and have been successfully used at a national or state level where many observations are available and sampling variation averages out \cite{abbott2020estimating,flaxman2020estimating,rtlive2020}. However, successful reopening will require programs which track spread at the level of particular colleges, workplaces, or towns, where \textit{partial observability} poses several challenges. \textit{First}, only a small number of infected people may be tested. It is estimated that only about 10\% of SARS-COV-2 infections in the US result in a confirmed test \cite{havers2020seroprevalence} and we could expect even fewer in populations with a high prevalence of asymptomatic or mild infections (e.g., college students). \textit{Second}, the biological properties of the test play an important role. For example, a PCR test which detects viral RNA will show positive results at different times than an antibody or antigen test. Further, there can be substantial heterogeneity across individuals. \textit{Third}, testing programs may collect samples in a particular way which impacts the observations. For example, one suggestion for schools and workplaces to reopen is to institute regular surveillance testing of a fraction of the population in order to detect outbreaks and catch asymptomatic carriers \cite{larremore2020test}. The observations will depend on the fraction of the population enrolled in testing (potentially small due to budget constraints) along with the sampling design (e.g., cross-sectional vs longitudinal).
	
\begin{figure}
		\centering
		\includegraphics[width=3.2in]{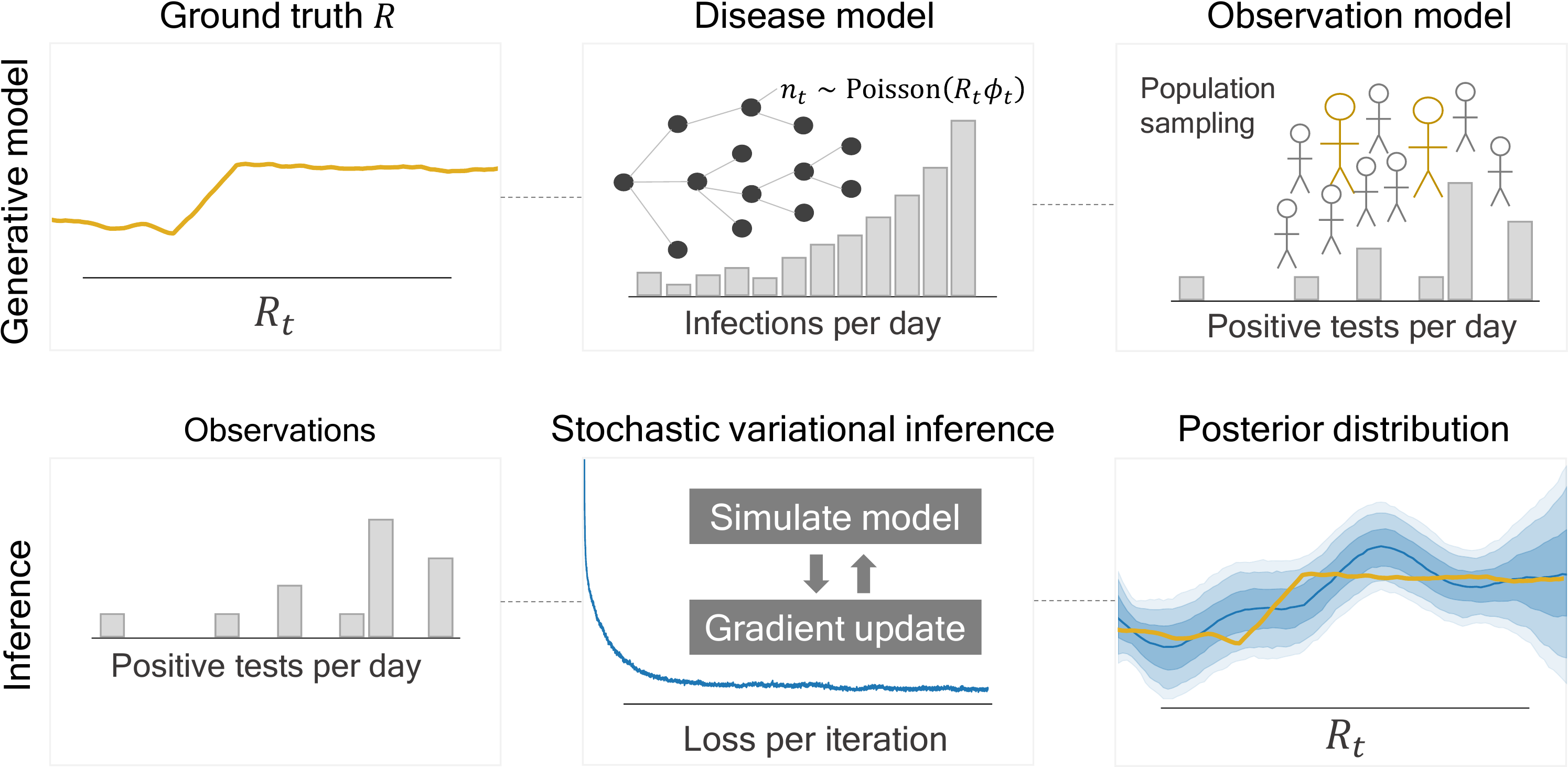}
		\caption{Illustration of our GPRt method. Top row: the generative model GPRt posits for the observed data. Bottom row: the inference process to recover a posterior over $R_t$.}
		\label{fig:illustration}
	\end{figure}

	This paper presents \textit{GPRt}, a novel Bayesian approach to estimating $R_t$ which accounts for partial observability in a flexible and principled manner (illustrated in Figure \ref{fig:illustration}). This method yields well-calibrated probabilistic estimates (the posterior distribution). Our model places a Gaussian process (GP) prior over $R_t$, allowing it to be an arbitrary smooth function. Then, we explicitly model the sampling process which generates the observations from the true trajectory of infections. While this substantially improves accuracy (as we show experimentally) it creates a much more difficult inference problem than has been previously considered. Specifically, our model contains tens or hundreds of thousands of discrete latent variables, preventing the application of out-of-the-box methods. Moreover, the values of many variables are tightly correlated in the posterior distribution, further complicating inference. To make inference computationally tractable, we propose a novel stochastic variational inference method, enabled by a custom stochastic gradient estimator for the variational objective. Extensive experiments show that our method recovers an accurate and well-calibrated posterior distribution in challenging situations where previous methods fail. 
	
	\section{Related work}
	
	There is a substantial body of work which attempts to infer unknowns in a disease outbreak. A frequent target for inference is the \emph{basic} reproduction number $R_0$ \cite{majumder2020early,riou2020pattern,wildermodeling}; by contrast, we attempt the more challenging task of estimating a reproduction number which can vary arbitrarily over time. There is a literature of both classic methods for estimating $R_t$ \cite{wallinga2004different,cori2013new,campbell2019bayesian} and newer methods developed for the COVID-19 pandemic and implemented in popular dashboards \cite{abbott2020estimating,rtlive2020}. None of these methods incorporate partial observability and we empirically demonstrate that GPRt improves substantially over methods in each category. Another strand of literature develops maximum likelihood or particle filter estimates of the parameters of an epidemiological model \cite{king2008inapparent,dureau2013capturing,cazelles2018accounting}. However, their work focuses on accommodating a complex model of the underlying disease dynamics; by contrast, we develop methods for probabilistically well-grounded inferences under a complex observation model. There is also a great deal of computational work more broadly concerned with disease control. Examples include optimization problems related to vaccination or quarantine decisions \cite{saha2015approximation,zhang2014dava,zhang2015controlling}, machine learning methods for forecasting (without recovering a probabilistic view of $R_t$) \cite{chakraborty2014forecasting,rekatsinas2015sourceseer}, and agent-based simulations of disease dynamics \cite{barrett2008interaction}. Our work complements this literature by allowing inference of a distribution over $R_t$ from noisy data, which can serve as the input that parameterizes an optimization problem or simulation.

	\section{Model}
	
	We now introduce a model for a disease process with a time-varying reproduction number. Subsequently, we introduce example models for how the observations are generated from the disease process which our framework can support. 
	\subsection{Disease model} 
	We use a standard stochastic model of disease transmission similar to other $R_t$ estimation methods \cite{wallinga2004different,cori2013new,campbell2019bayesian}; our contribution is a more powerful inference methodology which can accommodate complex observation models alongside a GP prior. Let $\bm{R} = [R_1...R_T]$ be the vector with $R_t$ for each time. From $\bm{R}$, the disease model defines a distribution over a vector $\bm{n} = [n_1...n_T]$ with the number of people newly infected each day. The main idea is that, over the course of a given individual's infection, they cause a Poisson-distributed number of new infections with mean determined by $\bm{R}$. Specifically, if on day $t$ individual $i$ has been infected for $h^i$ days the expected number of new infections caused by $i$ that day is
	\begin{align*}
	\lambda_t^i = w_{h^i} R_t.
	\end{align*}
	Here $w_{h^i}$ gives the level of infectiousness of an individual $h^i$ days post-infection. $w$ is normalized so that $\sum_h w_h = 1$. For later convenience, we will define $\phi_t  = \sum_{i = 1}^N w_{h^i}$ to be the total infectiousness in the population before scaling by $R_t$ ($N$ is the total population size). 
	
	
	
	Each day, each infected individual $i$ draws $n_t^i \sim \text{Poisson}(\lambda_t^i)$ other individuals to infect. We also incorporate infections from outside the population, with a mean of $\gamma$ such infections per day. We assume the rate of external infection is constant with respect to our time but our model could be extended to a time-varying $\gamma$. We treat $\gamma$ mostly as a nuisance parameter: our true objective is to infer $\bm{R}$, but doing so requires accounting for the potential that some detected cases are due to infections from outside the population. We define $n_t = \sum_{i = 1}^N n_t^i + \text{Poisson}(\gamma)$ to be the total number of new infections. Since $\bm{\phi}$ is fixed given the time series $\bm{n}$, we denote it as a function $\bm{\phi(n)}$. We denote the probabilistic disease model induced by a specific choice of $\bm{R}$ and $\gamma$ as $M(\bm{R}, \gamma)$ and the draw of a time series of infections from the model as $\bm{n} \sim M(\bm{R}, \gamma)$. 
	
	\subsection{Observation model}
	
	We now depart from the standard disease model used in previous work and describe a wide-ranging set of examples for how our framework can accommodate models of the process which generates the observed data from the latent (unknown) true infections. Previous work assumes either perfect observability or else the simplest of the three observation models we describe below (uniform undersampling). Our focus is where observations are generated by a medical test which confirms the presence of the pathogen of interest. Individuals have some probability of being tested at different times (depending on the testing policy adopted) and then test positive with a probability which depends on the biological characteristics of the disease and the test in question. 
	
	\subsubsection{Modeling tests}
	The kind of test employed determines when an individual is likely to test positive during the course of infection. For example, for COVID-19, PCR tests are commonly used to detect SARS-COV-2 RNA. They are highly sensitive and can detect early infections. Most infected individuals become PCR-negative within the week or two following infection as viral RNA is cleared \cite{kucirka2020variation}. By contrast, serological tests detect the antibodies produced after infection. An individual is not likely to test serologically positive until a week or more post-infection, but may then continue to test serologically positive for months afterwards \cite{iyer2020dynamics}. The observable data generated by a serological testing program is likely quite different than a PCR testing program since the time-frame and variance of when individuals test positive differs strongly. A range of other examples are possible (e.g., antigen tests) and can be easily incorporated into our framework. 
	
	Our model adopts a generic representation of a particular test as a distribution $D$ over $t_{\text{convert}}$, the number of days post-infection when an individual begins to test positive and $t_{\text{revert}}$, the number of days post-infection when they cease to test positive. For an infected individual $i$, we write $t^i_{\text{convert}}, t^i_{\text{revert}} \sim D$. Our method only assumes the ability to sample from $D$, meaning that we can directly plug in the results of lab studies assessing the properties of a test. $t^i_{\text{convert}}, t^i_{\text{revert}}$ are unobserved: we only get to see if an individual tests positive at a given point in time, not the full range of times that they \textit{would} have tested positive. 
	
	Next, we describe a series of example models for how and when individuals are tested, which reveal observations depending on the status ($t^i_{\text{convert}}, t^i_{\text{revert}}$) for each person who is tested. For convenience, we let $t^i_{\text{convert}} = \infty$ for an individual who is never infected. Note however that we can model false negatives by having $D$ sometimes set $t^i_{\text{convert}} = \infty$ for an infected individual, or false positives by returning finite $t^i_{\text{convert}}$ for an uninfected individual. We denote the number of observed positive tests on day $t$ as $x_t$. An observation model is a distribution over $\bm{x}$ given $\bm{n}$, denoted $\bm{x} \sim Obs(\bm{n})$. Each setting below describes one such distribution.
	
	\subsubsection{Uniform undersampling}
	
	In this setting, each individual who is infected is tested independently with some probability $p_{\text{test}}$ (e.g., if they individually decide whether to seek a test). To model this process, we introduce two new sets of latent random variables. First, a binary variable $z^i \sim \text{Bernoulli}(p_{\text{test}})$, indicating whether individual $i$ is tested.  Second, a delay $c^i$, giving the number of days between $t^i_{\text{convert}}$ and when individual $i$ is tested. We can integrate out the $z^i$ and obtain the following conditional distribution for the observed number of positive tests $x_t$:
	\begin{align*}
	x_t|c, t_{\text{convert}} \sim \text{Binomial}\left(\sum_{i = 1}^N 1[t = t^i_{\text{convert}} + c^i], p_{\text{test}}\right)
	\end{align*}
	where $1[\cdot]$ denotes the indicator function of an event. However, we cannot analytically integrate out $t_{\text{convert}}$ and $c$.
	
	\subsubsection{Cross-sectional testing}
	
	Here, a uniformly random sample of $s_t$ individuals are tested on each day $t$. This models a random screening program (e.g., testing random employees each day as they enter a workplace). In this case, we have
	\fontsize{9.5}{10}
	\begin{align*}
	x_t|t_{\text{convert}}, t_{\text{revert}} \sim \text{Binomial}\left(s_t, \frac{1}{N} \sum_{i = 1}^N 1[t^i_{\text{convert}} \leq t < t^i_{\text{revert}}]\right)
	\end{align*}
	\normalsize
	This expression provides the likelihood of $\bm{x}$ after conditioning on the latent variables $t^i_{\text{convert}}, t^i_{\text{revert}}$, though there is no closed-form expression conditioning only on $\bm{n}$.
	
	\subsubsection{Longitudinal testing}
	
	In this setting, a single sample from the population is chosen up front and every individual in the sample is tested every $d$ days. We again denote the total number of individual tested on day $t$ as $s_t$, but note that now the group of individuals who are tested repeats every $d$ days. Longitudinal testing offers different (and potentially more revealing) information than cross-sectional testing since when an individual first tests positive, we know that they did \textit{not} test positive $d$ days ago. However, it complicates inference by introducing correlations between the test results at different time steps. 
	Let $A_t$ denote the set of individuals who are tested at time $t$. We assume that the complete sample $\bigcup_{t = 1}^d A_t$ is chosen uniformly at random from the population, with the chosen individuals then randomly partitioned between the $d$ days. We have
	\begin{align*}
	x_t = \sum_{i \in A_t} 1[t^i_{\text{convert}} > t - d \text{ and } t^i_{\text{convert}} \leq t \text{ and } t^i_{\text{revert}} > t]
	\end{align*}
	where $t^i_{\text{convert}} > t - d$ captures that $i$ was not positive on their previous test. This introduces correlations between $x_t$ and $x_{t - d}$, so there is not a simple closed-form expression for the distribution of the time series $\bm{x}$ even after conditioning on $t^i_{\text{convert}}$ and $ t^i_{\text{revert}}$. (as there is in the cross-sectional case). We will instead build a flexible framework for inference which can just as well use a kind of sample of the log-likelihood.


	
	
	
	\section*{Inference problem}
	
	We will place a Gaussian process (GP) prior over $\bm{R}$, resulting in the following generative model:
	\begin{align*} 
	\bm{R} \sim \mathcal{GP}(&1, \mathcal{K}), \gamma \sim \text{Exp}(\bar{\gamma})\\
	&\bm{n} \sim M(\bm{R}, \gamma)\\
	&\bm{x} \sim Obs(\bm{n}).
	\end{align*}
	where $\mathcal{GP}(1, \mathcal{K})$ denotes a Gaussian process with constant mean 1 and kernel $\mathcal{K}$ and $\text{Exp}(\bar{\gamma})$ is an exponential prior on $\gamma$ with mean $\bar{\gamma}$. Given the observation $\bm{x}$, our goal is to compute the resulting posterior distribution over $\bm{R}$ and $\gamma$. However, is complicated by the fact that $\bm{x}$ is determined by a large number of discrete latent variables, primarily $\bm{n}$ (the time series of infections) and $\{t^i_{\text{convert}}, t^i_{\text{revert}}\}_{i=1}^N$, the times when each individual tests positive. A typical strategy for inference in complex Bayesian models is Markov Chain Monte Carlo (MCMC). However, MCMC is difficult to apply because of tight correlations between the values of variables over time: due to the GP prior, we expect values of $\bm{R}$ to be closely correlated between timesteps, and successive values of $\bm{n}$ are highly correlated via the model $M$. Formulating good proposal distributions for high-dimensional, tightly correlated random variables is notoriously difficult and has presented problems for GP inference via MCMC in other domains \cite{titsias2008markov}. 
	
	The other main approach to Bayesian inference is \emph{variational inference}, where we attempt to find the best approximation to the posterior distribution within some restricted family. Modern variational inference methods, typically intended for deep models such as variational autoencoders \cite{kingma2013auto}, use a combination of autodifferentiation frameworks and the reparameterization trick to differentiate through the variational objective \cite{kucukelbir2017automatic}. This process is highly effective for models with only continuous latent variables. However, our model has many thousands of discrete latent variables which cannot be reparameterized in a differentiable manner. Typical solutions to this problem would be to either integrate out the discrete variables or to replace them with a continuous relaxation \cite{jang2016categorical,vahdat2018dvae++}. Neither solution is attractive in our case -- the structure of the model does not allow us to integrate out the discrete variables analytically, while a continuous relaxation is infeasible because our latent variables have a strict integer interpretation (every infection requires in a particular individual becoming test-positive at particular points in time). 
	
	The last resort to differentiate through discrete probabilistic models is the score function estimator \cite{paisley2012variational}, which is often difficult to apply due to high variance. GPRt uses a combination of techniques which exploit the structure of infectious disease models to develop an estimator with controlled variance. \textit{First}, we develop a more tractable variational lower bound which is amenable to stochastic optimization. \textit{Second}, we hybridize the reparameterization and score function estimators across different parts of the generative model to take advantage of the properties of each component. \textit{Third}, we develop techniques to sample low-variance estimates of the log-likelihood for each of the observation models introduced earlier. These techniques are introduced in the next section.
	
	\section*{GPRt: variational inference algorithm}
	
	We now derive \emph{GPRt}, a novel variational inference method for $R_t$ estimation. GPRt approximates the true (uncomputable) posterior over $(\bm{R}, \gamma)$ via a multivariate normal distribution with mean $\bm{\mu}$ and covariance matrix $\Sigma$. $\mu_t$ is the posterior mean for $R_t$ while $\Sigma_{t, t'}$ gives the posterior covariance between $R_t$ and $R_{t'}$. $\mu_\gamma$ is the mean for $\gamma$ and $\Sigma_{\gamma, \cdot}$ gives its covariance with $R$. The diagonal $\Sigma_{t,t}$ gives the variance of the posterior over $R$ at each time, capturing the overall level of uncertainty. The aim is to find a $\bm{\mu}$ and $\Sigma$ which closely approximate the true posterior. Let $q(\bm{R}, \gamma|\bm{\mu}, \Sigma)$ denote the variational distribution. Let $p$ be the true generative distribution, where $p(\bm{R}, \gamma, \bm{x})$ is the joint distribution over $\bm{x}$ and $(\bm{R}, \gamma)$, $p(\bm{R}, \gamma)$ is the prior over $(\bm{R}, \gamma)$, and $p(\bm{R}, \gamma|\bm{x})$ is the posterior over $(\bm{R}, \gamma)$ after conditioning on $\bm{x}$.
	
	The aim of variational inference is to maximize a lower bound on the total log-probability of the evidence $\bm{x}$:
	\begin{align*}
	\log p(\bm{x}) &\geq  \E_{\bm{R}, \gamma \sim q}\left[\log p(\bm{R}, \gamma)\right] + \E_{\bm{R}, \gamma \sim q}\left[\log p(\bm{x}|\bm{R}, \gamma)\right]\\ &\quad\quad\quad\quad\quad\quad\quad\quad- \E_{\bm{R}, \gamma \sim q}\left[{\log q(\bm{R}, \gamma|\bm{\mu}, \Sigma)}\right]
	\end{align*}
	where the right-hand side is referred to as the \textit{Evidence Lower Bound (ELBO)}. Our goal is to maximize the ELBO via gradient ascent on the parameters $\bm{\mu}$ and $\Sigma$. This requires us to develop an estimator for the gradient of each term in the ELBO. The first term is the negative cross-entropy between $q$ and the prior $p(\bm{R}, \gamma)$. Because both $q$ and $p$ have simple parametric forms, this can be easily computed and differentiated. The last term is the entropy, which is similarly tractable. The middle term is the expected log-likelihood. Developing an estimator for the gradient of this term is substantially more complicated and will be our focus. In fact, for computational tractability we will actually develop an estimator for a lower bound on the expected log-likelihood; substituting this lower bound into the ELBO still gives a valid lower bound on $\log p(\bm{x})$ and so is a sensible objective.


	\subsection{Gradient estimator}
	
	The essential problem is that computing the log-likelihood of $\bm{R}$ requires integrating out the discrete latent variable $\bm{n}$ induced by the disease spread model, which is computationally intractable. The aim of this section is to develop the following stochastic estimator:
	
	\begin{theorem}
		Let $L$ be the Cholesky factor of $\Sigma$, $\xi \sim N(0, I)$, and $\begin{bmatrix}\bm{R} \\ \gamma\end{bmatrix} = \bm{\mu} + L \xi$. Let $\bm{n} \sim M(\bm{R}, \gamma)$. Finally, define
		\begin{align*}
		\hat{\nabla} = \nabla_{\bm{\mu}, L} \log M(\bm{n} | \bm{R}, \gamma) \log p(\bm{x}|\bm{n}).
		\end{align*}
		There exists a function $g(\bm{\mu}, \Sigma)$ with $\E_{\bm{R}, \gamma \sim q}\left[\log p(\bm{x}|\bm{R}, \gamma)\right]$ $\geq g(\bm{\mu}, \Sigma) \,\,\forall \bm{\mu}, \Sigma$ and $\E[\hat{\nabla}] = \nabla g(\bm{\mu}, \Sigma)$.\label{theorem:grad}
	\end{theorem} 
	
	Essentially, Theorem \ref{theorem:grad} states that $\hat{\nabla}$ is an unbiased estimator for a lower bound on the expected log-likelihood, exactly what we need to optimize a lower bound on $\log p(\bm{x})$ by stochastic gradient methods. Moreoever, as we will highlight below, we can efficiently compute the terms of $\hat{\nabla}$ via a combination of leveraging the structure of the disease model to apply autodifferentiation tools and novel sampling methods for the observation model. We now derive $\hat{\nabla}$.
	
	\begin{proof}
		Expanding the dependence of $\bm{x}$ on $\bm{n}$ we can rewrite the log-likelihood as \fontsize{9}{10}
		\begin{align*}
		\E_{\bm{R}, \gamma \sim q}\left[\log p(\bm{x}|\bm{R}, \gamma)\right]  = \E_{\bm{R}, \gamma\sim q(\bm{\mu}, \Sigma)}\left[\log \left(\E_{\bm{n} \sim M(\bm{R}, \gamma) }\left[p(\bm{x}|\bm{n})\right]\right)\right]
		\end{align*}\normalsize
		It is not clear how to develop a well-behaved gradient estimator for this expression because we wish to differentiate with respect to the parameters governing two nested expectations, one within the log. However, via Jensen's inequality, we can derive the lower bound \fontsize{9}{10}
		\begin{align*} 
		\E_{\bm{R}, \gamma\sim q}\left[\log p(\bm{x}|\bm{R}, \gamma)\right]  \geq \E_{\bm{R}, \gamma\sim q(\bm{\mu}, \Sigma)}\left[\E_{\bm{n} \sim M(\bm{R}, \gamma)}\left[\log p(\bm{x}|\bm{n})\right]\right],
		\end{align*}\normalsize
		pushing the log inside the expectation. We will substitute this bound into the ELBO, obtaining a valid lower bound to maximize. The key advantage is that our new lower bound admits an efficient stochastic gradient estimator. We start with the inner expectation and attempt to compute a gradient with respect to $\bm{R}$ (which controls the distribution of the simulation results $\bm{n}$). Using \emph{score function estimator} gives
		\begin{align*}
		&\nabla_{\bm{R}, \gamma} \E_{\bm{n} \sim M(\bm{R}, \gamma)}\left[\log p(\bm{x}|\bm{n})\right]\\ &= \E_{\bm{n} \sim M(\bm{R}, \gamma)}\left[\nabla_{\bm{R}, \gamma} \log M(\bm{n} | \bm{R}, \gamma) \log p(\bm{x}|\bm{n})\right]
		\end{align*}
		which expresses the gradient with respect to $(\bm{R}, \gamma)$ in terms of the gradient of the probability density of the disease model $M$ with respect to $(\bm{R}, \gamma)$. It turns out that $\log M(\bm{n}, \bm{\phi}|\bm{R}, \gamma)$ can be easily computed. Recall that $n_t = \sum_{i = 1}^N n_t^i + \text{Poisson}(\gamma)$, where $n_t^i \sim \text{Poisson}(R_t \phi_t^i)$. Using the Poisson superposition theorem, we have that $n_t \sim \text{Poisson}(\sum_{i = 1}^N R_t \phi_t^i + \gamma)$ (while $\phi_t$ is a deterministic function of $n_1...n_{t-1}$). Accordingly, we have that 
		\begin{align*}
		&\log M(\bm{n} | \bm{R}, \gamma) = \sum_{t = 1}^T \log \text{Pr}[n_t|n_1...n_T]\\
		&= \sum_{t = 1}^T n_t \log \left(R_t \phi_t(\bm{n}) + \gamma\right) - e \left(R_t \phi_t(n) + \gamma\right) - \log n_t!
		\end{align*}
		where the second line substitutes the Poisson log-likelihood. This expression can be easily differentiated with respect to $\bm{R}$ and $\gamma$ in closed form. Accordingly, we obtain an unbiased estimate of the gradient of our lower bound by sampling $\bm{n} \sim M(\bm{R}, \gamma)$ and computing 
		\begin{align*}
		\nabla_{\bm{R}, \gamma} \log M(\bm{n} | \bm{R}, \gamma) \log p(\bm{x}|\bm{n}).
		\end{align*}
		
		This suffices to estimate the gradient with respect to $(\bm{R}, \gamma)$. However, our goal is to differentiate with respect to $\bm{\mu}$ and $\Sigma$, which control the distribution over $(\bm{R}, \gamma)$. Fortunately, $\bm{R}$ and $\gamma$ are continuous. So, we can exploit the reparameterization trick by writing $(\bm{R}, \gamma)$ as a function of a random variable whose distribution is fixed. Specifically, since $\Sigma$ is positive semi-definite, it has a Cholesky decomposition $\Sigma = LL^T$ (for convenience, we actually optimize over $L$ instead of $\Sigma$). Sampling a standard normal variable $\xi \sim N(0, I)$ and letting $\begin{bmatrix}\bm{R} \\ \gamma\end{bmatrix} = \bm{\mu} + L \xi$ is equivalent to sampling $\bm{R},\gamma \sim N(\bm{\mu}, \Sigma)$. We rewrite the lower bound as 
		\begin{align*}
		g(\bm{\mu}, \Sigma) = E_{\xi \sim N(0, I)}\left[ \E_{\bm{n} \sim M(\bm{R}(\xi), \gamma(\xi))}\left[\log p(\bm{x}|\bm{n})\right]\right]
		\end{align*}
		where $\bm{\mu}$ and $L$ appear only as parameters of the deterministic function expressing $\bm{R}$ and $\gamma$ in terms of $\xi$, instead of in the distribution of a random variable. Taking a sample from each of the expectations and substituting the score function estimator now gives the desired expression for $\hat{\nabla}$.   \end{proof}
	
	Using Theorem \ref{theorem:grad}, our final gradient estimator will sample $b$ values for $\xi$, run the model $M$ once for each of the resulting values of $\bm{R}$ to sample $\bm{n}$, and then compute 
	\begin{align*}
	\frac{1}{b}\sum_{k = 1}^b  \nabla_{\bm{\bm{\mu}}, L} \log M(\bm{n}(k) |\bm{R}(\xi(k)), \gamma(\xi(k))) \log p(\bm{x}|\bm{n}(k)),
	\end{align*}
	easily accomplished with standard autograd tools given the closed-form expressions for $\bm{R}(\xi)$, $\gamma(\xi)$, and $\log M(\bm{n} | \bm{R})$. In practice, we also use the mean $\log p(\bm{x}|\bm{n}(k))$ as a simple control variate to reduce variance \cite{sutton1998introduction}.
	
	\newcommand{\ra}[1]{\renewcommand{\arraystretch}{#1}}
	\begin{table*}[h!]\centering
		\ra{1.3}
		\fontsize{6.5}{6.5}\selectfont
		\begin{tabular}{@{}rccccccccc@{}}\toprule
			\multicolumn{10}{c}{\textbf{Outbreak setting}}\\
			\midrule
			& \multicolumn{4}{c}{PCR} & & \multicolumn{4}{c}{Serological} \\
			\cmidrule{2-5} \cmidrule{7-10} 
			Longitudinal & 0.5\% & 1\% & 2\% & 5\% && 0.5\% & 1\% & 2\% & 5\%\\ \midrule 
			WT&0.481 $\pm$ 0.147&0.405 $\pm$ 0.11&0.395 $\pm$ 0.0947&0.401 $\pm$ 0.113&&0.445 $\pm$ 0.153&0.415 $\pm$ 0.117&0.423 $\pm$ 0.103&0.438 $\pm$ 0.147\\
			Cori&1.74 $\pm$ 0.774&1.18 $\pm$ 0.573&0.806 $\pm$ 0.373&0.546 $\pm$ 0.212&&1.55 $\pm$ 0.757&0.97 $\pm$ 0.577&0.621 $\pm$ 0.362&0.455 $\pm$ 0.191\\
			EpiNow&0.329 $\pm$ 0.211&0.265 $\pm$ 0.145&0.25 $\pm$ 0.135&0.267 $\pm$ 0.168&&0.308 $\pm$ 0.226&0.25 $\pm$ 0.171&0.232 $\pm$ 0.136&0.225 $\pm$ 0.134\\
			GPRt&\textbf{0.228 $\pm$ 0.0713}&\textbf{0.2 $\pm$ 0.055}&\textbf{0.183 $\pm$ 0.0579}&\textbf{0.186 $\pm$ 0.0692}&&\textbf{0.237 $\pm$ 0.0805}&\textbf{0.232 $\pm$ 0.0667}&\textbf{0.218 $\pm$ 0.0669}&\textbf{0.216 $\pm$ 0.0712}\\
			\midrule
			Cross-sectional & 0.05\% & 0.1\% & 0.2\% & 0.5\% && 0.05\% & 0.1\% & 0.2\% & 0.5\%\\  \midrule 
			WT&0.474 $\pm$ 0.149&0.396 $\pm$ 0.132&0.369 $\pm$ 0.102&0.358 $\pm$ 0.101&&0.472 $\pm$ 0.136&0.484 $\pm$ 0.135&0.501 $\pm$ 0.132&0.509 $\pm$ 0.123\\
			Cori&1.3 $\pm$ 0.676&0.859 $\pm$ 0.442&0.554 $\pm$ 0.184&0.502 $\pm$ 0.197&&1.12 $\pm$ 0.51&0.825 $\pm$ 0.363&0.664 $\pm$ 0.246&0.584 $\pm$ 0.167\\
			EpiNow&0.306 $\pm$ 0.199&0.277 $\pm$ 0.174&0.294 $\pm$ 0.184&0.302 $\pm$ 0.205&&\textbf{0.25 $\pm$ 0.146}&0.29 $\pm$ 0.181&0.269 $\pm$ 0.153&0.295 $\pm$ 0.174\\
			GPRt&\textbf{0.215 $\pm$ 0.063}&\textbf{0.178 $\pm$ 0.0509}&\textbf{0.177 $\pm$ 0.049}&\textbf{0.172 $\pm$ 0.0471}&&0.262 $\pm$ 0.09&\textbf{0.265 $\pm$ 0.076}&\textbf{0.249 $\pm$ 0.0791}&\textbf{0.238 $\pm$ 0.0732}\\\midrule
			Uniform underreporting & 1\% & 2\% & 5\% & 10\% && 1\% & 2\% & 5\% & 10\%\\  \midrule 
			WT&0.395 $\pm$ 0.105&0.389 $\pm$ 0.106&0.377 $\pm$ 0.111&0.382 $\pm$ 0.104&&0.407 $\pm$ 0.0937&0.425 $\pm$ 0.114&0.429 $\pm$ 0.133&0.408 $\pm$ 0.139\\
			Cori&0.892 $\pm$ 0.552&0.614 $\pm$ 0.355&0.412 $\pm$ 0.162&0.38 $\pm$ 0.108&&0.98 $\pm$ 0.553&0.587 $\pm$ 0.245&0.431 $\pm$ 0.149&0.38 $\pm$ 0.139\\
			EpiNow&0.311 $\pm$ 0.193&0.31 $\pm$ 0.186&0.359 $\pm$ 0.231&0.394 $\pm$ 0.245&&\textbf{0.254 $\pm$ 0.136}&\textbf{0.212 $\pm$ 0.128}&0.251 $\pm$ 0.139&0.267 $\pm$ 0.16\\
			GPRt&\textbf{0.204 $\pm$ 0.0806}&\textbf{0.22 $\pm$ 0.0878}&\textbf{0.181 $\pm$ 0.0677}&\textbf{0.181 $\pm$ 0.0467}&&0.26 $\pm$ 0.0948&0.259 $\pm$ 0.0839&\textbf{0.222 $\pm$ 0.0865}&\textbf{0.233 $\pm$ 0.0807}\\\midrule
			\multicolumn{10}{c}{\textbf{Random trend setting}}\\
			\midrule
			& \multicolumn{4}{c}{PCR} & & \multicolumn{4}{c}{Serological} \\
			\cmidrule{2-5} \cmidrule{7-10} 
			Longitudinal & 0.5\% & 1\% & 2\% & 5\% && 0.5\% & 1\% & 2\% & 5\%\\ \midrule 
			
			WT&0.427 $\pm$ 0.149&0.345 $\pm$ 0.101&0.321 $\pm$ 0.101&0.292 $\pm$ 0.104&&0.398 $\pm$ 0.118&0.322 $\pm$ 0.0851&0.288 $\pm$ 0.0937&0.26 $\pm$ 0.0646\\
			Cori&1.28 $\pm$ 0.678&0.872 $\pm$ 0.512&0.622 $\pm$ 0.326&0.392 $\pm$ 0.159&&1.04 $\pm$ 0.666&0.57 $\pm$ 0.325&0.358 $\pm$ 0.151&0.28 $\pm$ 0.0768\\
			EpiNow&0.332 $\pm$ 0.233&0.321 $\pm$ 0.195&0.337 $\pm$ 0.232&0.349 $\pm$ 0.244&&0.364 $\pm$ 0.225&0.291 $\pm$ 0.207&0.304 $\pm$ 0.185&0.296 $\pm$ 0.209\\
			GPRt&\textbf{0.199 $\pm$ 0.0745}&\textbf{0.187 $\pm$ 0.0652}&\textbf{0.181 $\pm$ 0.0551}&\textbf{0.157 $\pm$ 0.0476}&&\textbf{0.232 $\pm$ 0.0733}&\textbf{0.216 $\pm$ 0.0754}&\textbf{0.213 $\pm$ 0.0699}&\textbf{0.194 $\pm$ 0.0694}\\\midrule
			Cross-sectional & 0.05\% & 0.1\% & 0.2\% & 0.5\% && 0.05\% & 0.1\% & 0.2\% & 0.5\%\\  \midrule 
			
			WT&0.392 $\pm$ 0.123&0.335 $\pm$ 0.107&0.319 $\pm$ 0.101&0.284 $\pm$ 0.096&&0.381 $\pm$ 0.128&0.404 $\pm$ 0.118&0.406 $\pm$ 0.13&0.396 $\pm$ 0.115\\
			Cori&0.94 $\pm$ 0.535&0.581 $\pm$ 0.217&0.478 $\pm$ 0.159&0.411 $\pm$ 0.117&&0.68 $\pm$ 0.242&0.61 $\pm$ 0.219&0.513 $\pm$ 0.137&0.485 $\pm$ 0.117\\
			EpiNow&0.359 $\pm$ 0.202&0.356 $\pm$ 0.191&0.421 $\pm$ 0.225&0.383 $\pm$ 0.215&&0.362 $\pm$ 0.242&0.438 $\pm$ 0.216&0.436 $\pm$ 0.265&0.456 $\pm$ 0.264\\
			GPRt&\textbf{0.192 $\pm$ 0.068}&\textbf{0.182 $\pm$ 0.0641}&\textbf{0.168 $\pm$ 0.0512}&\textbf{0.149 $\pm$ 0.0467}&&\textbf{0.242 $\pm$ 0.0889}&\textbf{0.246 $\pm$ 0.0877}&\textbf{0.233 $\pm$ 0.0925}&\textbf{0.221 $\pm$ 0.0788}\\
			\midrule
			Uniform underreporting & 1\% & 2\% & 5\% & 10\% && 1\% & 2\% & 5\% & 10\%\\  \midrule 
			WT&0.285 $\pm$ 0.095&0.275 $\pm$ 0.0884&0.267 $\pm$ 0.107&0.259 $\pm$ 0.109&&0.316 $\pm$ 0.0943&0.3 $\pm$ 0.0917&0.287 $\pm$ 0.0892&0.282 $\pm$ 0.0908\\
			Cori&0.558 $\pm$ 0.368&0.396 $\pm$ 0.187&0.326 $\pm$ 0.128&0.281 $\pm$ 0.109&&0.527 $\pm$ 0.298&0.382 $\pm$ 0.205&0.303 $\pm$ 0.108&0.276 $\pm$ 0.105\\
			EpiNow&0.348 $\pm$ 0.244&0.315 $\pm$ 0.179&0.383 $\pm$ 0.238&0.336 $\pm$ 0.201&&0.308 $\pm$ 0.173&0.356 $\pm$ 0.236&0.318 $\pm$ 0.22&0.349 $\pm$ 0.237\\
			GPRt&\textbf{0.172 $\pm$ 0.0694}&\textbf{0.17 $\pm$ 0.0632}&\textbf{0.163 $\pm$ 0.071}&\textbf{0.181 $\pm$ 0.0682}&&\textbf{0.2 $\pm$ 0.0793}&\textbf{0.213 $\pm$ 0.0831}&\textbf{0.206 $\pm$ 0.0848}&\textbf{0.211 $\pm$ 0.0742}\\\bottomrule
		\end{tabular}
		
		\caption{Mean absolute error of each method averaged over instances and time points for each setting, along with standard deviation of the absolute error. ``PCR" and ``Serological" denote settings where the observations are generated by the respective testing method. Individual column headings give the percentage of the population enrolled in testing. } \label{table:mae}
	\end{table*}

	\subsection{Computing the likelihood}
	
	We now turn to the task of computing the log-likelihood function $\log p(\bm{x}|\bm{n})$, which measures the log-likelihood of observing the sequence of positive test results $\bm{x}$ given $\bm{n}$ new infections per day. Unfortunately, the log-likelihood is not available in closed form for any of the settings that we consider because it depends on additional latent variables (e.g., $t_{\text{convert}}, t_{\text{revert}},c,$ or $A$). We will show that it suffices to develop an estimator which lower-bounds the log-likelihood and that such estimators can be efficiently implemented for each of the observation models we consider. Specifically, denote the collection of latent variables used in a particular observation model as $\alpha$. Then, we have 
	\begin{align*}
	\log p(\bm{x}|\bm{n}) = \log \left(\E_{\alpha}[p(\bm{x}|\bm{n}, \alpha) | \bm{n}]\right),
	\end{align*}
	which presents a similar difficulty as in developing our earlier lower bound: sampling $\alpha$ to approximate the inner expectation does not result in an unbiased estimator due to the outer log. Using Jensen's inequality in the same way gives
	\begin{align*}
	\log p(\bm{x}|\bm{n}) \geq \E_{\alpha}\left[\log p(\bm{x}|\bm{n}, \alpha) | \bm{n}\right],
	\end{align*}
	and so substituting the right-hand side into our variational objective preserves validity of the lower bound. The RHS has the crucial advantage that we can now develop an unbiased estimator by drawing a single sample of $\alpha$, which can then be substituted into the stochastic gradient estimator of Theorem \ref{theorem:grad}. That is, for each of the simulation results $n(1)...n(b)$ we sample a corresponding value for the latent variables, $\alpha(1)...\alpha(b)$ and use the gradient estimator\fontsize{9}{10}
	\begin{align*}
	\frac{1}{b}\sum_{k = 1}^b  \nabla_{\bm{\mu}, L} \log M(\bm{n}(k) | \bm{R}(\xi(k)), \gamma(\xi(k))) \log p(\bm{x}|\bm{n}(k), \alpha(k))
	\end{align*}\normalsize
	This works without issue for the uniform undersampling and cross-sectional models where we can obtain a closed form for the log likelihood after conditioning on the appropriate latent variables. However, the longitudinal testing model presents additional complications. In particular, after sampling the latent variables $t_{\text{convert}}$, $t_{\text{revert}}$, and $A_t$, the number of positive tests becomes deterministic quantity. Denote this simulated trajectory of positive tests $\tilde{x}$. If $\tilde{x} = x$, then $p(\bm{x}|$ $t_{\text{convert}}$, $t_{\text{revert}}, A) = 1$ and otherwise $p(\bm{x}|$ $t_{\text{convert}}$, $t_{\text{revert}}, A) = 0$. This renders the above gradient estimator useless because $\log p(\bm{x}|\bm{n}(k), \alpha(k)) = -\infty$ unless the simulated trajectory \emph{exactly} matches the observed data (a very low-probability event). While $-\infty$ is technically a valid lower bound for the variational objective, it is not very useful for optimization. Essentially, we need to develop a lower-variance estimator where the lower bound is more useful. 
	
	We now present one such improved estimator. The intuition is that we can marginalize out a great deal of the randomness in the naive estimator by only revealing the results of random draws determining $A_t$ a single individual at a time. We start by sampling $t_{\text{convert}}$ and $t_{\text{revert}}$. Note that we can expand $\log p(\bm{x}|\bm{n}, t_{\text{convert}}$, $t_{\text{revert}}) = \sum_{t = 1}^T \log p(x_t|x_1...x_{t-1}, \bm{n}, t_{\text{convert}}$, $t_{\text{revert}})$ and consider the likelihood at each day $t$ after conditioning on the results observed on previous days. To compute an estimate for this sum, we introduce a new object, the series of matrices $C^t$. At each time $t$, $C^t[t_1, t_2]$ denotes the number of individuals who have $t_{\text{convert}} = t_1$, $t_{\text{revert}} = t_2$, and have not yet actually tested positive by time $t$. Since $A_t$ is selected uniformly at random from the population, independent of the infection process, the $x_t$ individuals who test positive on day $t$ are drawn uniformly at random from the set of all individuals who converted between days $t - d$ and $t$, and who have not yet reverted. Let $n_{\text{draws}}$ denote the number of individuals in $A_t$ who have not yet tested positive by time $t$ and $n_{\text{conv}} = \sum_{t_1 = t - d}^{t} \sum_{t_2 = t+1}^T C^t[t_1, t_2]$ denote the number of individuals who are ``eligible" to test positive at time $t$. Now $x_t|x_1...x_{t-1}, \bm{n}, C^t$ follows a binomial distribution with  $n_{\text{draws}}$ draws and success probability $\frac{n_{\text{conv}}}{N - \sum_{i = 1}^{t-1} x_i}$. Accordingly, the log-likelihood $\log p(x_t|x_1...x_{t-1}, \bm{n}, C^t)$ can be computed in closed form. After this, we can sample $C^t | C^{t+1}$ by selecting a uniformly random individual to remove from $C^{t+1}$. We can view this as iteratively revealing the test-positive members of $A_t$ after conditioning on the sequence of previous test results, instead of sampling the entire set up front as in the naive method.

	\section{Experimental results}

	\begin{figure*}
		\centering
		\includegraphics[width=1.3in]{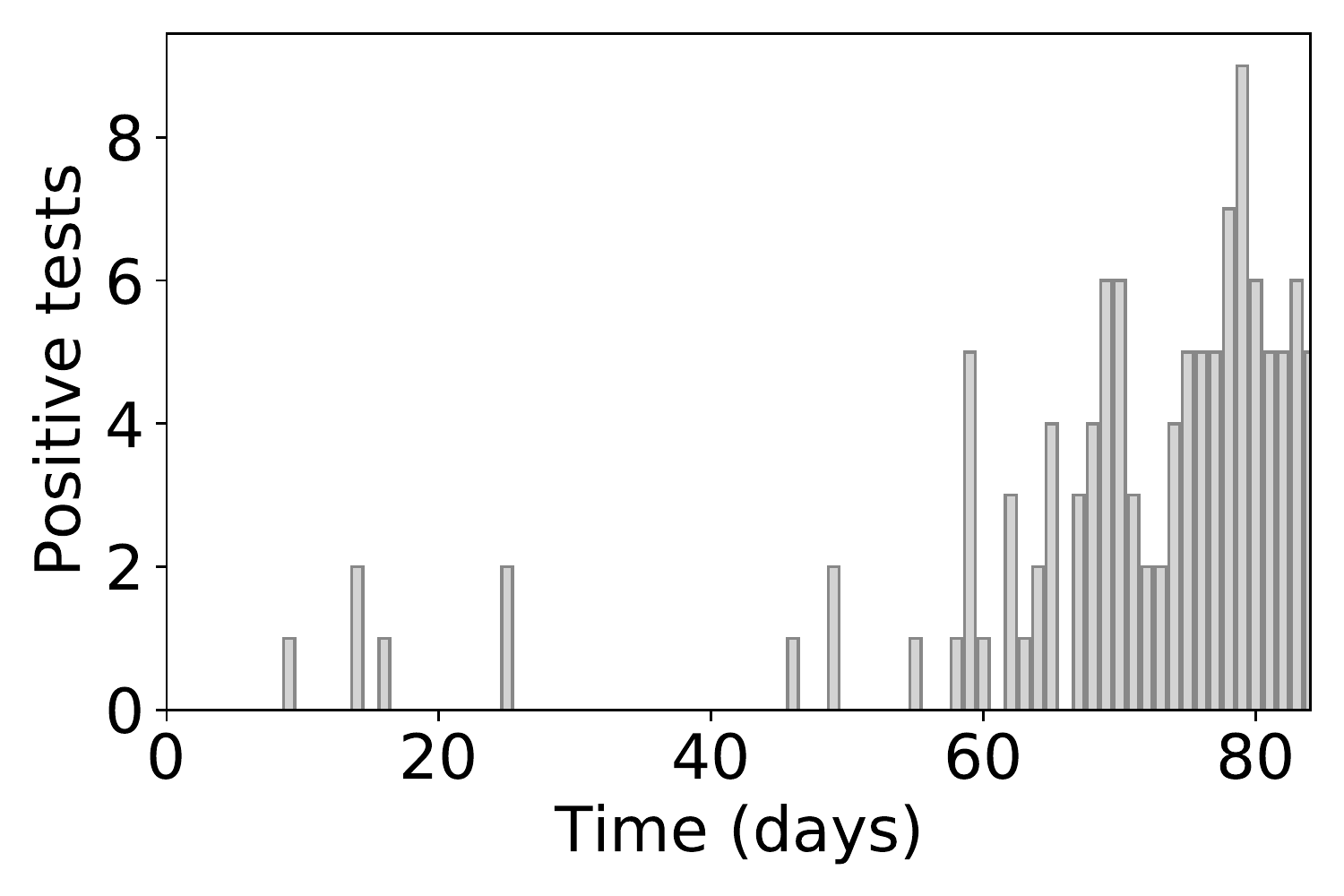}
		\includegraphics[width=1.3in]{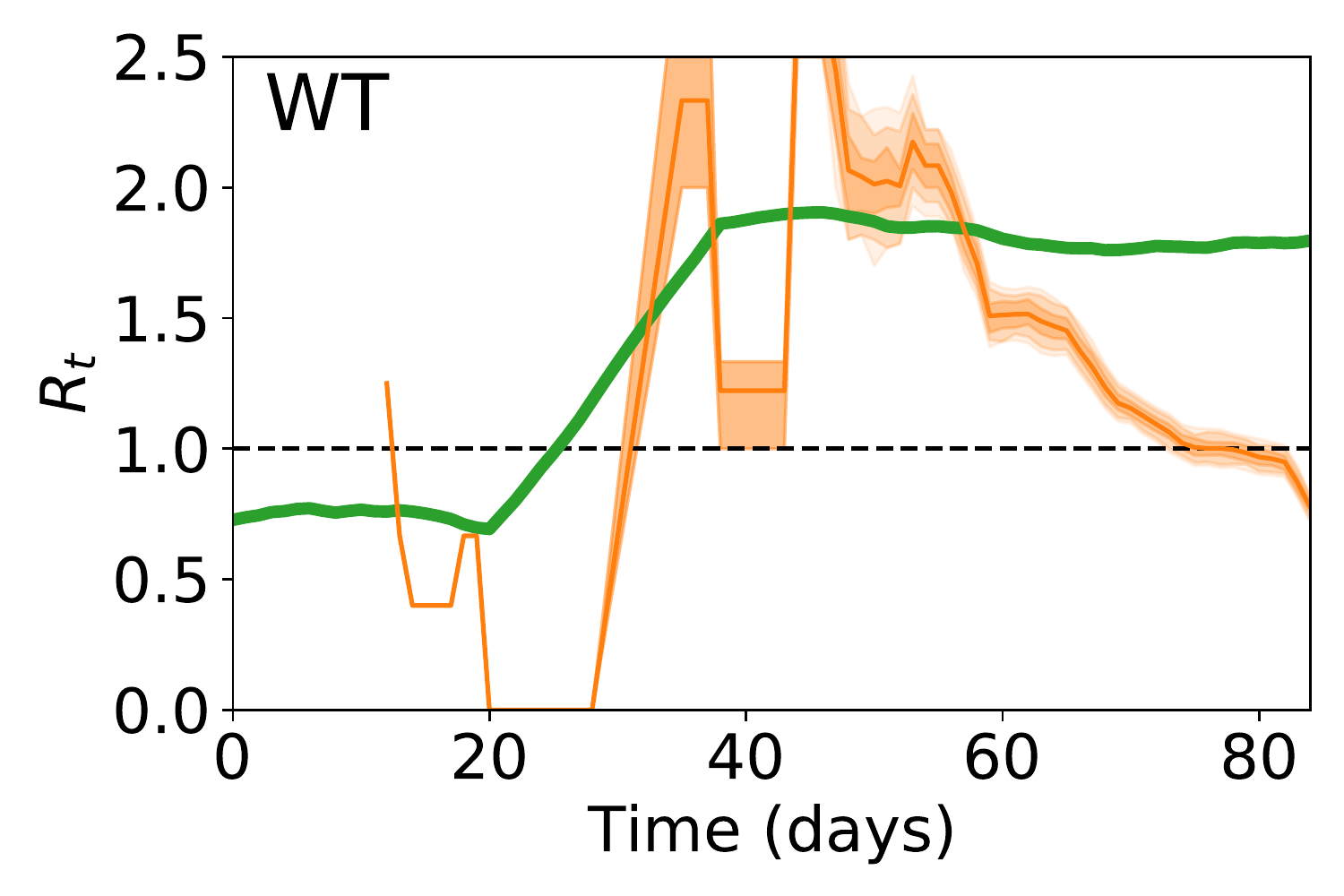}
		\includegraphics[width=1.3in]{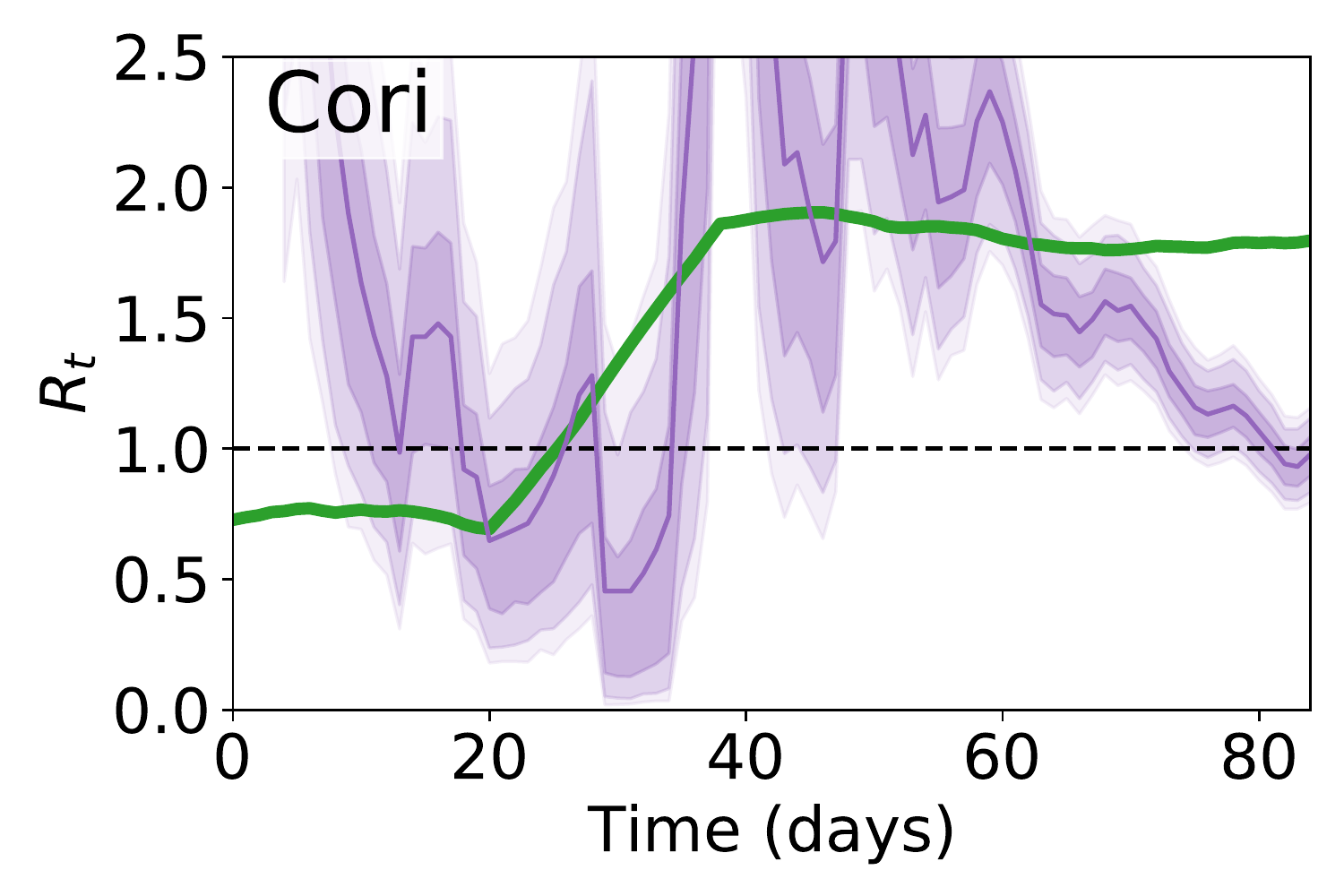}
		\includegraphics[width=1.3in]{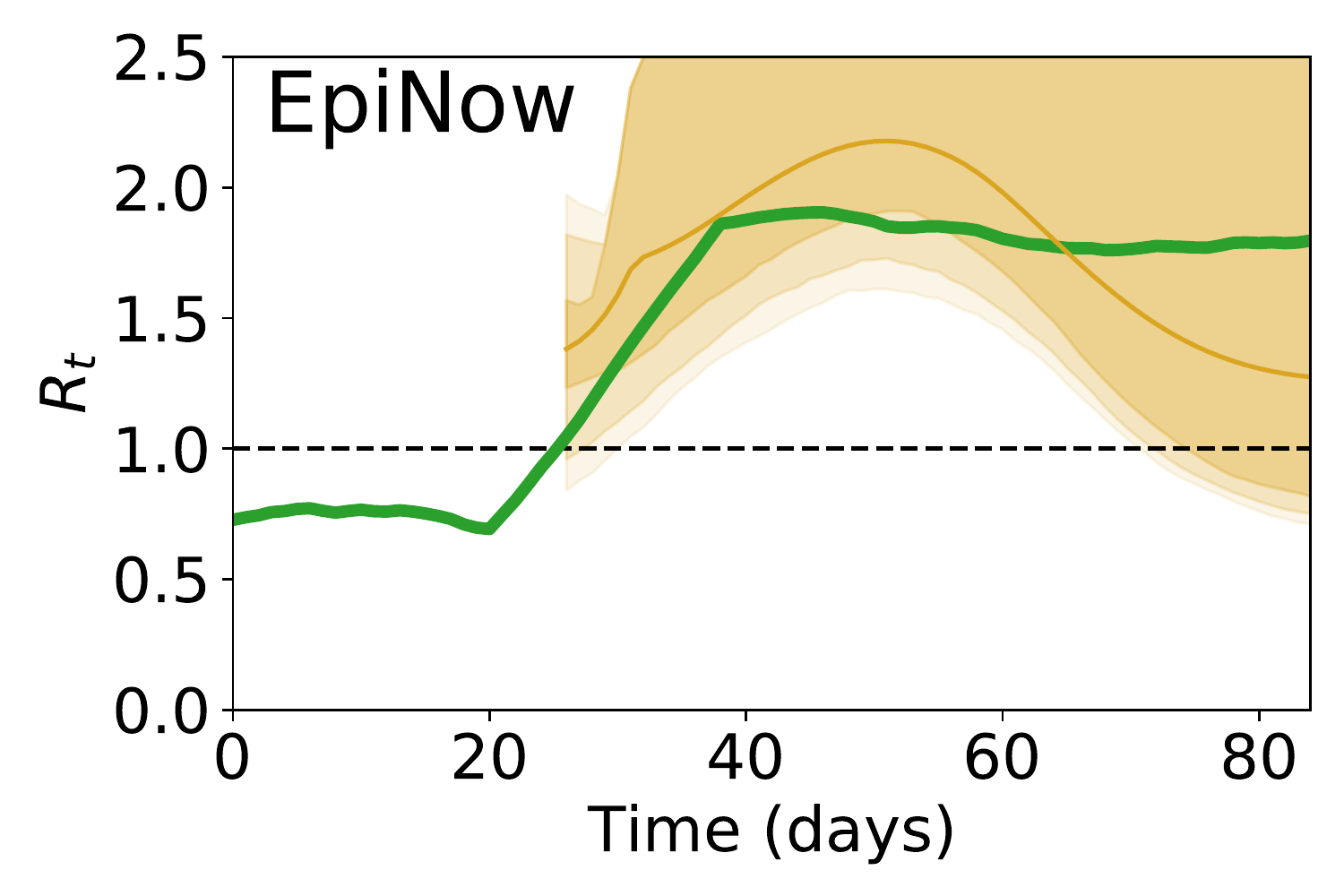}
		\includegraphics[width=1.3in]{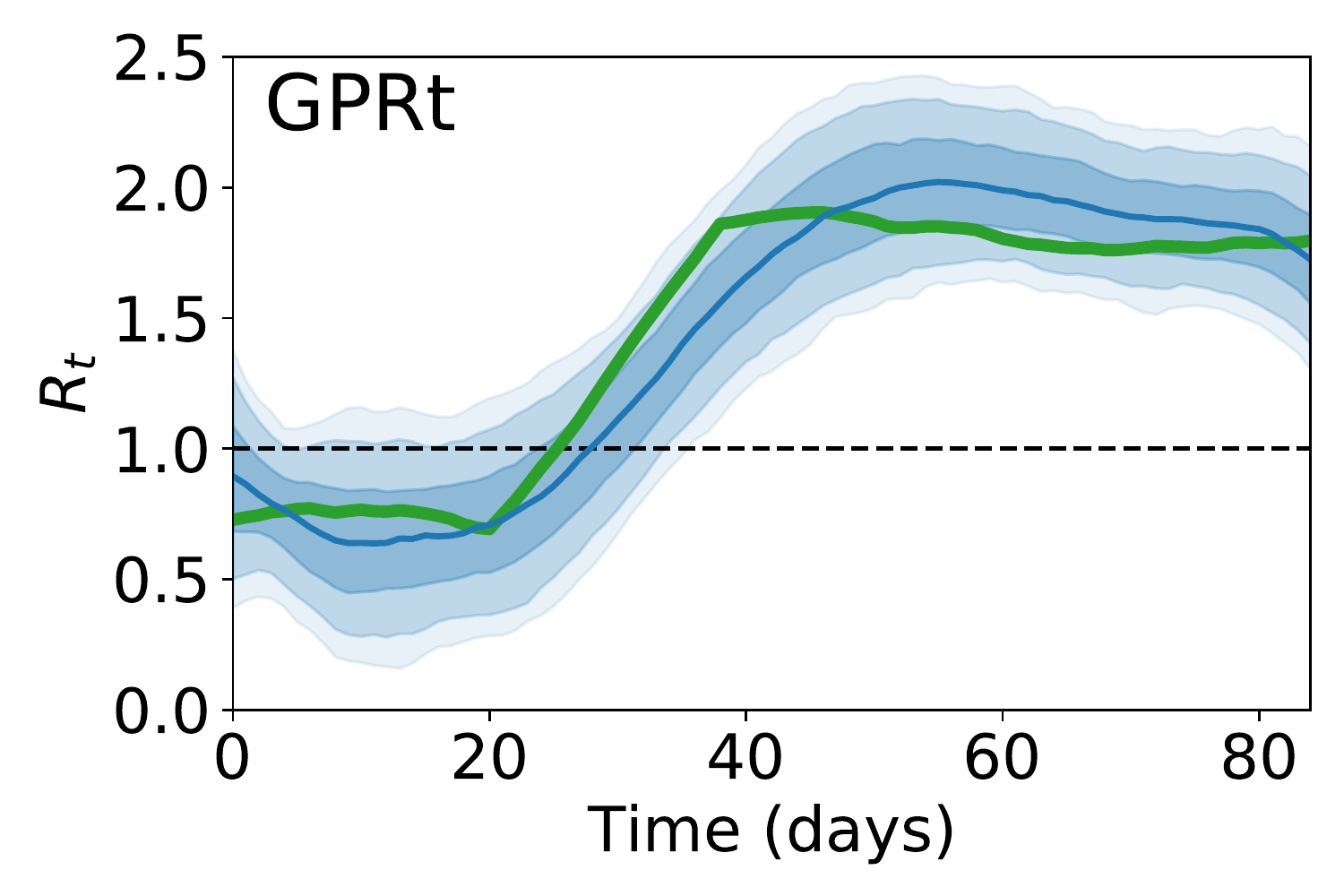}
		\caption{Observed $x_t$ and the distribution over $R_t$ returned by each method on an example in the outbreak setting with longitudinal sampling, $d = 14$, and a 1\% sample. The green line gives the ground truth $R_t$.}
		\label{fig:example}
	\end{figure*}
	
	\begin{figure}
		\centering
		\includegraphics[width=3.2in]{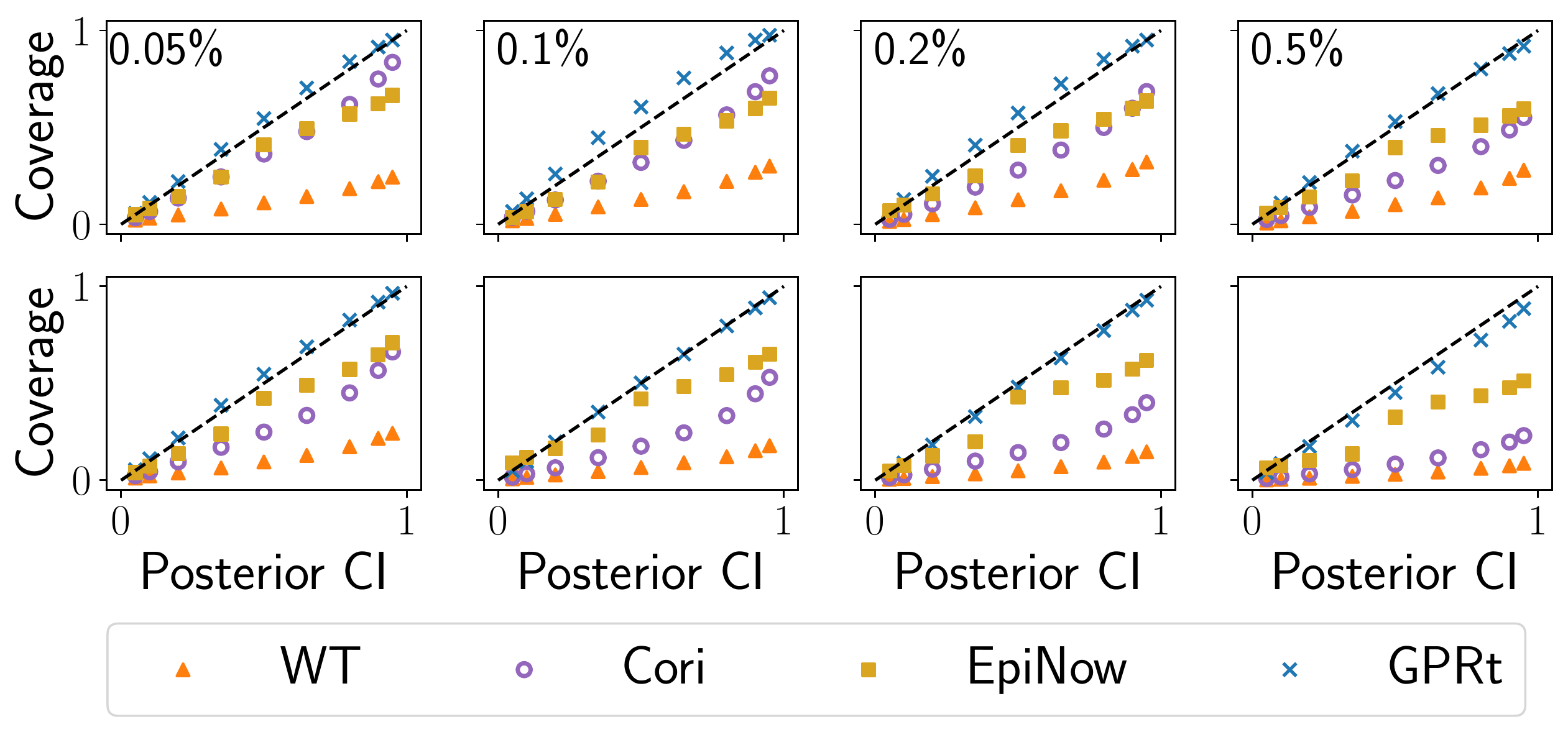}
		\caption{Calibration of each method for cross-sectional testing. Top row: PCR. Bottom row: serological. The number in the upper-left hand corner of each column gives $\%$ tested per day. Each individual plot shows the calibration of each method for that setting. Each $(x, y)$ point gives the fraction of the ground truth data points ($y$) which are covered by the method's posterior interval at level $\bm{x}$. So, e.g., a point placed at (0.6, 0.7) would indicate that 60\% of the ground truth data points fell into the method's 70\% credible interval. The dashed diagonal line shows perfect calibration and points lying closer to this line indicate better calibration. }
		\label{fig:calibration}
	\end{figure}
	
	
	We test the performance GPRt vs standard baselines on a wide variety of settings. We choose three baselines which have been recommended by leading epidemiologists as methods of choice for COVID-19 \cite{gostic2020practical}. First is the \emph{Wallinga-Teunis (WT)} method \cite{wallinga2004different}, which uses the distribution of the time between an infected person and their secondary infections to simulate possible who-infected-who scenarios, each of which induces a particular $R_t$. WT assumes that cases are observed exactly and that there is no delay in observation. Second is the method of \emph{Cori et al.\ (Cori)} \cite{thompson2019improved,cori2013new} which computes a Bayesian posterior distribution in a similar Poisson branching process infection model. The difference is that their method does not place a GP prior over $R_t$ (instead the posterior factorizes over time) and does not model the sampling method or delays. For both WT and Cori, we apply a common heuristic to correct for time-to-reporting delays, which is shift the method's predictions by the mean delay. Third is \emph{EpiNow} \cite{abbott2020estimating}, a MCMC method recently developed for COVID-19 which places a GP prior over $\bm{R}$ and accounts for the delay distribution, but does not model partial observability. 
	
	We test each method in an array of settings, with different distributions for both the true value of $R_t$ and the observations. We include two different settings for the ground truth $R_t$. First, the \textit{outbreak} setting where $\bm{R}$ starts below 1 and rises above 1 at a random time. Second, the \textit{random trend} setting where $\bm{R}$ follows a linear trend which changes randomly at multiple points in time. Details of the settings and other experimental parameters are in the appendix.
	
	
	We also include different observation models characterized by the test used, the sampling method, and the sample size. We include both \emph{PCR} and \emph{serological} tests, using previously estimated distributions for $D$ \cite{kucirka2020variation,iyer2020dynamics}. We also include three sampling models introduced earlier: \emph{uniform underreporting}, \emph{cross-sectional}, and \emph{longitudinal}. Finally, for each of the four combinations of tests and sampling method, we include four different sample sizes. Many sizes model a challenging setting with sparse observations, representing highly limited testing capacity. Note that the sample sizes evaluated are different for each method because they have different interpretations, e.g. 1\% in the cross-sectional case means sampling 1\% of the population each day while in the longitudinal case it would mean 1\% every $d$ days. For each setting, Table \ref{table:mae} shows the mean absolute error between the posterior mean $\bm{R}$ produced by each method and the ground truth. Each entry averages over 100 instances. For longitudinal testing we use $d = 14$; results for other values are very similar (see appendix).
	
	
	\textit{Across almost all settings, our method has lower MAE than any baseline, often by a substantial margin (reducing error by a factor of 2-10x).} Notably, GPRt performs well even with extremely limited data (e.g., when testing 0.05\% of the population per day or when 1\% of infections are observed). Performance improves with more data, but the gains limited (e.g., 0.02-0.04 MAE), indicating that our method is able to make effective use of even very sparse data.  
	
	Figure \ref{fig:example} shows a representative example. The observed data is quite sparse, with 0-8 positive tests observed per day. Our method recovers a posterior which closely tracks the ground truth. WT produces an estimate which is correlated with the ground truth but has many fluctuations and overly tight confident intervals. Cori is not appropriate for data this sparse and produces a widely fluctuating posterior. EpiNow does not return an estimate for much of the time series, only estimating the part with denser observations (we gave the baselines an advantage by only evaluating their MAE where they returned an estimate). Moreoever, even in the higher-observation portion, it is less accurate than GPRt.
	
	Finally, Figure \ref{fig:calibration} shows \emph{calibration}, a metric which evaluates the entire posterior (not just the mean). Intuitively, calibration reflects that, e.g., 90\% of the data should fall into the 90\%-credible interval of the posterior. Calibration is critical for the posterior distribution to be interpretable as a valid probabilistic inference, and for it to be useful in downstream decision making. Figure \ref{fig:calibration} shows the fraction of the data which is covered by the credible intervals of each method. This figure shows cross-sectional testing in the outbreak setting, but results for other settings are very similar (see appendix). \textit{GPRt is close to perfectly calibrated (the dotted diagonal line) while the baseline methods are not well calibrated}. The baselines suffer from two problems. First, as to be expected from their higher MAE, they are biased and so their credible intervals often exclude the truth. Second, they are over-confident: paradoxically, their calibration worsens with increased data since the larger sample size makes them more confident in their erroneous prediction. We conclude that GPRt offers uniquely well-calibrated inferences.

	\bibliographystyle{aaai}
	\bibliography{bib}

\end{document}